
\magnification=1200
\def\nbody{{N--body}}
\vsize=9truein
\hsize=6.5truein
\hoffset=1truein
\voffset=1truein
\overfullrule=0pt
\centerline{A Test of the Adhesion Approximation for}
\smallskip
\centerline{Gravitational Clustering }
\vskip .25in
\centerline{{\bf Adrian L. Melott},$^1$ {\bf Sergei F. Shandarin},$^1$ and}
\centerline{{\bf David H. Weinberg}$^2$}
\medskip
\item{$^1$} Department of Physics and Astronomy, University of Kansas,
Lawrence, Kansas 66045
\item{$^2$} Institute for Advanced Study, Princeton, New Jersey 08540
\vskip .5in

\baselineskip=12pt
\noindent {\bf ABSTRACT}
\medskip
We quantitatively compare a particle implementation of the adhesion
approximation to fully non--linear, numerical ``\nbody" simulations.
Our primary
tool, cross--correlation of \nbody\ simulations with the adhesion
approximation,
indicates good agreement,
better than that found by the same test performed with the Zel'dovich
approximation (hereafter ZA). However, the cross--correlation is not as good as
that of the truncated Zel'dovich approximation (TZA),
obtained by applying the Zel'dovich approximation after smoothing the
initial density field with a Gaussian filter.
We confirm that the adhesion approximation produces an excessively
filamentary distribution.
Relative to the \nbody\ results, we also find that: (a) the power spectrum
obtained from the adhesion approximation is more accurate than that from ZA or
TZA, (b) the error in the phase angle of Fourier components is worse than that
from TZA, and (c) the mass distribution function is more accurate than
that from ZA or TZA.
It appears that adhesion performs well statistically,
but that TZA is more accurate dynamically, in the
sense of moving mass to the right place.
\vfill
\noindent {\bf Subject Heading}: Galaxies, formation, clustering--large--scale
structure of the Universe
\vfil\eject

\noindent {\bf I. INTRODUCTION}
\medskip
The large--scale structure of the Universe is thought to have arisen from
primarily gravitational processes acting to amplify primordial density
fluctuations.  In the limit of small fluctuations, the differential equations
for the density contrast can be solved by linear perturbation theory
(see, e.g., Peebles [1980]).
Zel'dovich (1970)
proposed an approximation in which the linear--order particle velocities
are extrapolated, instead of the linear density contrast.
It has often been supposed that this approximation is appropriate
only for initial fluctuation spectra with very
little power on small scales. However, Coles, Melott and Shandarin (1993;
hereafter CMS) compared a number of approximate schemes for gravitational
evolution, starting from initial conditions with power spectra
$P(k)\propto k^n$, and
found that the Zel'dovich approximation (hereafter ZA)
was the most successful of these for all indices $n\leq +1$.
\medskip
There is considerable benefit to applying the same tests to a series of
approximations, so that they can be compared with each other.
In this paper we report the results of applying the CMS tests to the
adhesion approximation, proposed by Gurbatov, Saichev and Shandarin (1985,
1989) and implemented in a particle code by Weinberg and Gunn (1990).
(For further discussion of the adhesion approximation, see
Kofman {\it et al.} [1992] and references therein).
\medskip
CMS proposed a simple extension of ZA, obtained by smoothing the initial
density field {\it before} applying ZA, in order to suppress the effect of
strongly non-linear modes, which tend to scatter particles out
of collapsed regions.  This ``truncated Zel'dovich approximation'' (hereafter
TZA) was considerably refined and tested by
Melott, Pellman and Shandarin (1993; hereafter MPS).
They found that the optimal way to smooth the initial conditions is
by convolution with a Gaussian filter
whose radius is a specific, spectrum--dependent multiple of the scale
of nonlinearity.
TZA is a major improvement over ZA, and it as
fast as one step in an \nbody\ simulation.
Any numerical implementation of ZA automatically truncates Fourier
modes below the resolution limit (e.g.\ the Nyquist frequency in a grid code),
but in TZA the truncation is handled in a controlled and optimized manner.
In this paper, we will compare the adhesion approximation (hereafter AA)
to \nbody\ results and to ZA and TZA.
\medskip
For a complete discussion of ZA, TZA, and AA, we refer the reader to the
papers cited above.  Here we limit
ourselves to a simple physical description of AA, which can itself be regarded
as an extension of ZA.
With an appropriate choice of variables, ZA can be
regarded as simple inertial motion, continuing the initial velocities of fluid
elements (see e.g. Shandarin \& Zeldovich, 1989).
The approximation works much better than one might think based on
this description, in large part because the potential has a larger
coherence length than the density if the effective spectral index is
$n\leq 1$. AA adds to the inertial motion a viscosity force
$\nu\cdot
\nabla^2 {\bf v}$, where ${\bf v}$ is the velocity and $\nu$ is the viscosity
coefficient. This viscosity mimics some of the effects of
nonlinear gravity, by eliminating the relative velocity of intersecting flows
and causing fluid elements to ``stick'' when they fall into caustics.
However, for finite $\nu$ this term is also nonzero in the voids,
where the flow accelerates away from underdense perturbations,
and in these regions it is likely to degrade the accuracy of the approximation.
\medskip
The addition of the viscosity term to inertial motion yields Burgers' equation.
If the initial velocity field is a potential flow (as expected in
gravitational instability models), then Burgers' equation admits an exact
integral expression for the velocity field at any later time.
This integral can be evaluated by steepest-descent in the limit
that the viscosity approaches (but does not equal) zero.  In this approach
(Gurbatov, Saichev and Shandarin 1985, 1989;
Nusser and Dekel, 1990; Kofman, Pogosyan and Shandarin 1992; Kofman {\it et
al.} 1992; Sahni, Sathyaprakash and Shandarin 1991),
the pancakes are infinitely thin, and the flow is exactly that given
by ZA outside of multistream regions.  From a geometrical analysis of the
velocity field, one can derive the skeleton of the structure (i.e.\ the
location of sheets, filaments, and knots) at any time, but not a detailed
distribution of matter inside collapsed regions.
\medskip
In this paper we use the particle implementation of AA described by
Weinberg and Gunn (1990).  In this method, one evaluates the Burgers
integral by Gaussian convolution, using a finite value of the viscosity
parameter.  The resulting code is closer in spirit to an \nbody\ code;
one integrates particle orbits, at each timestep using the velocity
field implied by the solution to Burgers' equation.
\bigskip
\noindent {\bf II. SIMULATIONS}
\medskip
In our AA simulations, we used the smallest value of the viscosity that
did not produce numerical overflows (see Weinberg and Gunn [1990] for
further discussion).
We checked the choice of timestep by comparing to similar runs with shorter
timesteps.
The adhesion simulations had initial conditions identical to one
realization set of \nbody\ simulations,
described in Melott and Shandarin (1993).
We used power law initial density fluctuation spectra, $P(k)\propto k^n$
for $n=-2,-1,0, +1$. We chose for analysis in all of these the moment when
rms fluctuations are just going nonlinear $(\delta\rho/\rho=1)$ at a
wavelength of $L/8$, where $L$ is the box size.
At this output time, nonlinear structures are well resolved, but the
scale of nonlinearity is small enough that the simulations' periodic boundary
conditions do not cause problems.
We also checked results at
a later stage when the nonlinear wavelength was $L/4$.
The \nbody\ and adhesion simulations both used
128$^3$ particles on a $128^3$ mesh.
\medskip
Figures 1a, 2a, 3a, and 4a  show slices one cell thick through the \nbody\
simulations for the four initial power spectra.
In these greyscale renderings, regions below the mean density are white, and
regions above a density contrast of 10 are black. Figures 1b to 4b show
corresponding slices from the adhesion simulations with the same
initial conditions. Figures 1c to 4c show results from TZA,
the most successful of the approximations previously tested in this series.
\medskip
As noted elsewhere (e.g.\ Weinberg and Gunn [1990]),
the adhesion simulations look more filamentary than the full
\nbody\ simulations. This is a reflection of the fact that their
``superpancakes" (see Melott and Shandarin 1993) are less broken up into
subcondensations. This difference in texture is more pronounced for larger
values of $n$.  Also, there seem
to be some condensations in the adhesion model that have no counterparts in
the \nbody\ run, though this mismatch may be an artifact of plotting thin
slices.  Although the TZA figures have fewer objects than either of the
others, their locations agree well with those of the primary condensations
in the \nbody\ run.
\vfill\eject
\noindent {\bf III. QUANTITATIVE COMPARISON}
\medskip
The most direct comparison we can make with \nbody\ asks whether the
adhesion simulations put mass in the same place.
To address this question quantitatively, we study the
cross--correlation of the adhesion models with the \nbody\ models in the manner
of CMS.  We define the cross-correlation statistic $S$ as
$$S={<\delta_1\; \delta_2>\over \sigma_1\; \sigma_2}~,\eqno (1)$$
where $\delta_1$
and $\delta_2$ are the local density contrast in the \nbody\ and adhesion
simulations at the same spot, and $\sigma_1$ and $\sigma_2$ are
the standard deviations of the two density fields.
For identical density fields, $S=1$.
When the fields are defined at very high resolution, small errors in the
precise positions of mass concentrations will destroy the correlation
between them.  We therefore compute $S$ for a variety of Gaussian smoothings,
which are applied in the same way to the two fields.  The heavy lines
in Figure 5 plot $S$ against the value of $\sigma$ in the smoothed
\nbody\ density field.  The light lines plot $S$ against $\sigma$ for TZA.
We see that:
\medskip
\item{(a)} The match to \nbody\ is worse for larger $n$, since initial
conditions with more small scale power have more strongly nonlinear modes,
which cannot be followed by the approximations.
\medskip
\item{(b)} By comparison with CMS, we find that AA crosscorrelates about as
well as ZA for $n=-2$, and better for all the other indices.
\item{(c)} In all cases, TZA performs better than AA on this test.
\medskip
Figure 6 compares power spectra of the \nbody\ simulations (heavy solid lines),
the adhesion simulations (light solid lines), and TZA (light dashed lines).
\nbody\ and AA are shown at two epochs,
when the nonlinear scale is $k_{nl}=L/8$ and $L/4$.
TZA is shown only at $k_{nl}=L/8$.
Small differences in the linear
(small--$k$) part of the spectrum appear to be a numerical artifact of the
\nbody\ code, since both approximations agree better with linear perturbation
theory in this regime.
The error may be related to the very low fluctuation amplitude
used in the initial
conditions --- one of us (DHW) has found similar behavior in a different
code when starting from very small initial fluctuations.
At its worst, it represents a 25\% error in power
after an expansion factor of about 5000.
\medskip
The adhesion approximation underestimates large--$k$ power, but it does a
better
job overall than any approximation tested so far. As we see from the dashed
lines, TZA has a considerably larger error in the nonlinear part of the
spectrum.
\medskip
Distributions are characterized by both amplitudes and phases for their Fourier
components. We tested for phase angle agreement by calculating
$<$cos $\theta$$>$,
where $\theta$ is the difference in phase angle of the corresponding Fourier
coefficients
and the averaging is over spherical shells in wavenumber. Figure 7 plots
$<$cos $\theta >$ against $k$. As expected, the agreement declines
steadily with increasing nonlinearity. The $n=+1$ model is
substantially worse than the others, with gradual improvement through to $n=-2$
as $n$ declines.  MPS find the same trends for TZA.  However, the phase
errors for TZA are {\it smaller} than those for AA.  It is probably these
smaller phase errors that account for TZA's higher cross--correlation with
\nbody.
\medskip
Figure 8 plots the mass density distribution function: $N(\rho)$ is the
number of cells with density in the range $\rho \rightarrow \rho + d\rho$.
For this test, the density fields are defined by cloud-in-cell weighting
of the $128^3$ particle distributions onto a $64^3$ mesh.
Both AA and TZA underestimate the number of high--density pixels and
overestimate the number of low-- and moderate--density ones, but AA is much
more successful here, as it is for the power spectrum.
\bigskip
\noindent {\bf IV. DISCUSSION}
\medskip
We summarize our conclusions and compare with previous work:
\smallskip
\item{(a)} The adhesion approximation is an substantial improvement over the
original Zel'dovich (1970) approximation in all aspects of its performance
(except, of course, computational speed).
\smallskip
\item{(b)} We measure AA's dynamical accuracy by cross--correlation with the
\nbody\ density fields.  The agreement is quite good, but not as good as that
found for the truncated Zel'dovich approximation by MPS.
\smallskip
\item{(c)} AA reproduces the \nbody\ power spectrum and mass density
distribution better than any other approximation that we have tested so far.
\smallskip
\item{(d)} AA makes greater errors than TZA in the phases of
Fourier coefficients of the mass distribution.
\smallskip
\item{(e)} By combining the above results with our visual examination of the
greyscale plots, we infer that AA is doing a reasonable job of making
condensations but is putting them in somewhat incorrect positions.
\smallskip
\item{(f)} In most respects, AA performs better than
the frozen--flow approximation -- see Melott, Lucchin, Matarrese,
and Moscardini (1993).
However, AA is much more computer intensive than frozen--flow.
\smallskip
\item{(g)} Of the methods that we have studied to date,
it appears that TZA is the best dynamical approximation, in the
sense of moving mass to the right place. AA is the best statistical
approximation, in that it comes closest to reproducing the statistical
results of \nbody\ simulations, at least for the $P(k)$ and $N(\rho)$
statistics that we have examined here.
\medskip
It is not at all clear what are the intrinsic sources of errors in the
adhesion particle method.  In one test case, we found that doubling the
value of the viscosity coefficient had almost no impact on the results,
which suggests that finite viscosity of the amplitude that we are
using here is not an important source of error.
Gurbatov {\it et al.} (1985, 1989)
applied the adhesion method in the limit of vanishing
viscosity, using it to find the skeleton of structure, not the details of the
mass distribution.  That implementation of AA cannot be compared to
\nbody\ simulations in the same way as the particle--pushing implementation
examined here, and it is not clear that it will make the same errors in
locating collapsed structures.  We think that further investigation of
this question is warranted, but it is outside the scope of the present study.
We should also note that we have compared AA and TZA for one stage only,
when $k_{nl}=8k_f$.  We found qualitatively similar results for $k_{nl}=4k_f$,
and we believe that our conclusions will hold for other stages as well.
\medskip
If one wants a ``poor man's \nbody'' method for evolving specified initial
conditions, TZA has clear advantages over the particle implementation of
AA: it is simpler, much faster, uses less memory, and produces more
accurate results (in terms of cross--correlation).
It has similar advantages over other approximations that we have tested.
Babul {\it et al.} (1993) have used adhesion to generate initial
conditions for simulations of the explosion scenario, and adhesion may
provide a useful computational technique for other specific applications.
However, the adhesion approximation will probably make
its most important contributions
as a tool for analytic calculations and as a source of physical
insight into the formation of large--scale structure.

\bigskip
\noindent {\bf ACKNOWLEDGEMENTS}
\medskip
Our computations were performed at the National Center for Supercomputing
Applications, Urbana, Illinois, USA. Research at the University of Kansas was
supported by NASA grant NAGW--2923 and NSF grants AST--9021414 and NSF EPSCoR
grant OSR--9255223.  DHW is supported by the W. M. Keck Foundation and
NSF grant PHY92-45317.
\bigskip
\noindent {\bf REFERENCES}
\bigskip
\def\ref{\par\noindent\hangindent\parindent\hangafter1}
\ref
Babul, A., Weinberg, D.H., Dekel, A. and Ostriker, J.P. 1993, ApJ, submitted
\ref
Coles, P., Melott, A.L and Shandarin, S.F. 1993, MNRAS 260, 765 (CMS)
\medskip
\ref
Gurbatov, S.N., Saichev, A.I. and Shandarin, S.F. 1985, Sov.Phys.Dokl. 30, 921
\medskip
\ref
Gurbatov, S.N., Saichev, A.I. and Shandarin, S.F. 1989, MNRAS 236, 385
\medskip
\ref
Kofman, L., Pogosyan, D. and Shandarin, S.F. 1990, MNRAS, 242, 200
\medskip
\ref
Kofman, L., Pogosyan, D., Shandarin, S.F. and Melott, A.L. 1992, ApJ 393, 437
\medskip
\ref
Melott, A.L. and Shandarin, S.F. 1993, ApJ 410, 469
\medskip
\ref
Melott, A.L., Pellman, T. and Shandarin, S.F. 1993, MNRAS, submitted (MPS)
\medskip
\ref
Melott, A.L., Lucchin, F., Matarrese, S., and Moscardini, L. 1993, MNRAS,
submitted
\medskip
\ref
Nusser, A. and Dekel, A. 1990, ApJ 363, 14
\medskip
\ref
Peebles, P.J.E. 1980, The Large--Scale Structure of the Universe, Princeton
Univ. Press, Princeton, NJ
\medskip
\ref
Sahni, V., Sathyaprakash, D. and Shandarin, S.F. 1993, in preparation
\medskip
\ref
Shandarin, S., and Zeldovich, Ya.B. 1989, Rev.Mod.Phys. 61, 185
\medskip
\ref
Weinberg, D.H. and Gunn, J.E. 1990, MNRAS 247, 260
\medskip
\ref
Zel'dovich, Ya.B. 1970, Astr.Astrophys. 5, 84
\vfill\eject
\noindent {\bf FIGURE CAPTIONS}
\medskip
\ref
{\bf Figure 1}: A greyscale plot of thin (L/128) slices through the
simulation cubes for $n=+1$ initial conditions, at the stage when
$k_{nl}=8k_f$. (a) The \nbody\ simulation. (b) The adhesion approximation (AA).
(c) The optimum, Gaussian--truncated Zel'dovich approximation (TZA).
All statistics below are calculated for this stage unless otherwise specified.
\medskip
\ref
{\bf Figure 2}: As in Figure 1, but for $n=0$ initial conditions.
\medskip
\ref
{\bf Figure 3}: As in Figure 1, but for $n=-1$ initial conditions.
\medskip
\ref
{\bf Figure 4}: As in Figure 1, but for $n=-2$ initial conditions.
\medskip
\ref
{\bf Figure 5}: The cross--correlation $S$ between the \nbody\ density
field and the density field of the approximate simulation (see equation [1]).
Bold lines plot the cross--correlation for AA against
the rms fluctuation of the \nbody\ density field,
after both fields are smoothed by convolution with identical
Gaussian windows of various sizes.
The initial power spectra are
$n=+1$ (longdash/shortdash), $n=0$ (shortdash), $n=-1$ (longdash), and
$n=-2$ (dotdash).  Lighter lines show the cross--correlation of TZA,
for the same spectra.
\medskip
\ref
{\bf Figure 6}:  Heavy solid lines show power spectra
of the evolved \nbody\ simulations at two stages
($k_{nl}=8k_f$ and $4k_f$).  Light solid lines show spectra of the
adhesion simulations at the same epochs.  Dashed lines show spectra
from TZA (shown at only one stage, $k_{nl}=8k_f$, to prevent confusion).
\medskip
\ref
{\bf Figure 7}: The average effective phase error in the adhesion simulations,
quantified by $<\cos\; \theta >$ as described in the text.  Different lines
represent
$n=+1$ (longdash/shortdash), $n=0$ (shortdash), $n=-1$ (longdash), and $n=-2$
(dotdash).
\medskip
\ref
{\bf Figure 8}: The mass density distribution function in the \nbody\
simulations
(heavy solid lines), AA (light solid lines), and TZA (dashed lines).
$N(\rho)$ is the number of cells with density (in units of the mean density)
in the range $\rho \rightarrow \rho + d\rho$.
\bye